\documentclass[
reprint,
superscriptaddress,
amsmath,amssymb,
aps,
prd,
floatfix,
]{revtex4-2}

\usepackage{graphicx}
\usepackage{dcolumn}
\usepackage{bm}
\usepackage[english]{babel}
\usepackage[utf8]{inputenc}
\usepackage[T1]{fontenc}
\usepackage{microtype}
\usepackage[table,dvipsnames]{xcolor}
\usepackage[
colorlinks=true,
breaklinks=true,
hyperfootnotes=true,
]{hyperref}
\usepackage{booktabs}
\usepackage{subfigure}
\usepackage{ulem}
\usepackage{multirow}
\usepackage{threeparttable}
\usepackage{makecell}
\begin{document}
	
	\preprint{APS/123-QED}
	
	\title{Rapid eccentric spin-aligned binary black hole waveform generation based on deep learning}
	
	\author{Ruijun Shi}
	\affiliation{School of Physics and Astronomy, Beijing Normal University, Beijing, 100875, China}
        \affiliation{Institute for Frontiers in Astronomy and Astrophysics, Beijing Normal University, Beijing, 102206, China}
	
	\author{Yue Zhou}
	\affiliation{Peng Cheng Laboratory, Shenzhen 518055, China}
	
	\author{Tianyu Zhao}
        \affiliation{Center for Gravitational Wave Experiment, National Microgravity Laboratory, Institute of Mechanics, Chinese Academy of Sciences, Beijing, 100190, China}
        
	\author{Zhixiang Ren}
 \altaffiliation{Corresponding author}
        \email{renzhx@pcl.ac.cn}
	\affiliation{Peng Cheng Laboratory, Shenzhen 518055, China}	
 
	\author{Zhoujian Cao}
 \altaffiliation{Corresponding author}
        \email{zjcao@bnu.edu.cn}
	\affiliation{School of Physics and Astronomy, Beijing Normal University, Beijing, 100875, China}
        \affiliation{Institute for Frontiers in Astronomy and Astrophysics, Beijing Normal University, Beijing, 102206, China}
        \affiliation{School of Fundamental Physics and Mathematical Sciences, Hangzhou Institute for Advanced Study, UCAS, Hangzhou, 310024, China}
	
	\date{\today}

\begin{abstract}
Accurate waveform templates of binary black holes (BBHs) with eccentric orbits are essential for the detection and precise parameter estimation of gravitational waves (GWs).
While \texttt{SEOBNRE} produces accurate time-domain waveforms for eccentric BBH systems, its generation speed remains a critical bottleneck in analyzing such systems. 
Accelerating template generation is crucial to data analysis improvement and valuable information extraction from observational data.
We present \texttt{SEOBNRE\_AIq5e2}, an innovative AI-based surrogate model that crafted to accelerate waveform generation for eccentric, spin-aligned BBH systems. 
\texttt{SEOBNRE\_AIq5e2} incorporates an advanced adaptive resampling technique during training, enabling the generation of eccentric BBH waveforms with mass ratios up to 5, eccentricities below 0.2, and spins $|\chi_z|$ up to 0.6. 
It achieves an impressive generation speed of 4.3 ms per waveform with a mean mismatch of $1.02 \times 10^{-3}$. 
With the exceptional accuracy and rapid performance, \texttt{SEOBNRE\_AIq5e2} emerges as a promising waveform template for future analysis of eccentric gravitational wave data.

\end{abstract}
\maketitle

\section{Introduction}
The detection of the first gravitational wave (GW), GW150914 \cite{Abbott2016, abbottBinaryBlackHole2016, abbottGW150914FirstResults2016}, has unveiled a new perspective on understanding the universe.
So far, GWTC-3 has released nearly one hundred compact binary coalesce (CBC) events \cite{abbottGWTC1GravitationalWaveTransient2019, abbottGWTC2CompactBinary2021, abbottGWTC3CompactBinary2023a}. 
Traditional GW data analysis has primarily focused on waveforms that are generated by binary black hole (BBH) systems in circular orbits. 
In ground-based GW detection, we usually assume that gravitational waves have been circularized due to the effect of gravitational radiation. This assumption is consistent with the current GW detection results \cite{ abbottGWTC1GravitationalWaveTransient2019, abbottGWTC2CompactBinary2021,abbottGWTC3CompactBinary2023a}. 
However, in recent years, there has been an increasing attention on detecting signs of eccentricity in present GW signals \cite{lowerMeasuringEccentricityBinary2018, abbottSearchEccentricBinary2019, romero-shawSearchingEccentricitySignatures2019a, nitzSearchEccentricBinary2020, romero-shawSignsEccentricityTwo2021, favataConstrainingOrbitalEccentricity2022a, boninoInferringEccentricityEvolution2023, xuMeasurabilityPrecessionEccentricity2023}.

Currently, the analysis of GW190521 provides evidence of eccentricity residuals \cite{abbottGW190521BinaryBlack2020, romero-shawGW190521OrbitalEccentricity2020, gayathriEccentricityEstimateBlack2022, iglesiasEccentricityEstimationFive2024}.
Eccentricity carries critical information about the astrophysical environments in which compact binaries form and evolve \cite{zevinEccentricBlackHole2019, cardosoEccentricityEvolutionCompact2021,zevinImplicationsEccentricObservations2021, kneeRosettaStoneEccentric2022, divyajyotiBlindSpotsBiases2024}. 
In certain extreme conditions, such as those found in dense stellar clusters \cite{zevinEccentricBlackHole2019}, near supermassive black holes \cite{tagawaEccentricBlackHole2021} and triple stars systems \cite{antoniniBinaryBlackHole2017}, binary black hole systems can retain significant orbital eccentricity as they enter the GW frequency bands. 
They are detectable by ground- and space-based GW detectors. 
Future space-based GW detection missions, including LISA \cite{amaro-seoaneLaserInterferometerSpace2017}, Taiji\cite{huTaijiProgramSpace2017}, and Tianqin\cite{Luo2016}, are expected to detect a broader range of GW signals. 
They have the potential to reveal a significant population of BBHs with eccentric orbits \cite{breivikDISTINGUISHINGFORMATIONCHANNELS2016, nishizawaConstrainingStellarBinary2017}. 
These observations will provide new insights into the formation channels and dynamic interactions that lead to eccentric compact binaries, deepening our understanding of the astrophysical processes in high-density environments \cite{zevinEccentricBlackHole2019}.

The GW detection and parameter estimation from GW detectors rely on a high-accuracy waveform model \cite{theligoscientificcollaborationGuideLIGOVirgoDetector2020}. 
Numerical relativity (NR) provides the most accurate waveform template, but NR requires a lot of computing resources and cannot be directly used in GW data analysis. 
The approximate templates used for data analysis are mainly effective-one-body (EOB) model \cite{taracchiniEffectiveonebodyModelBlackhole2014} and IMRPhenom model \cite{husaFrequencydomainGravitationalWaves2016, khanFrequencydomainGravitationalWaves2016}. 
And the widely used eccentricity BBH waveform templates include \texttt{SEOBNRE} \cite{caoWaveformModelEccentric2017,liuHighermultipoleGravitationalWaveform2022, liuEffectiveonebodyNumericalrelativityWaveform2024}, \texttt{SEOBNRv4EHM} \cite{khalilRadiationreactionForceMultipolar2021, ramos-buadesEffectiveonebodyMultipolarWaveforms2022}, \texttt{TEOBResumS-Dalí} \cite{chiaramelloFaithfulAnalyticalEffectiveonebody2020, albertiniEffectiveonebodyWaveformsExtrememassratio2024a, nagarEffectiveonebodyWaveformModel2024a},  \texttt{ENIGMA} \cite{huertaEccentricNonspinningInspiral2018a, chenObservationEccentricBinary2021}, \texttt{TaylorF2ECC} \cite{mooreGravitationalwavePhasingLoweccentricity2016, moore3PNFourierDomain2019}, \texttt{EccentricFD}, and \texttt{EccentricTD} \cite{tanayFrequencyTimedomainInspiral2016,  tiwariProposedSearchDetection2016, lalsuite, swiglal}.
Compared to circular orbit waveforms, the evolution of waveforms in eccentric orbits is more complex. \texttt{SEOBNRE} is one of the most precise eccentric GW waveform models with a precision adequate for GW data analysis. 
At total mass $M_t=60 M_\odot$ and $f_0=20\text{Hz}$, the CPU generation speed of \texttt{SEOBNRE} is 2.1s, the generation speed becomes the main bottleneck when about one million waveforms are required to estimate a single eccentric BBH.

The application of graphics processing units (GPUs) holds considerable promise for substantially improving computational efficiency in the field of gravitational astronomy \cite{katzGPUacceleratedMassiveBlack2020a, katzAssessingDataanalysisImpact2022, strubAcceleratingGlobalParameter2023}. 
The IMRPhomon models such as BBHx \cite{katzGPUacceleratedMassiveBlack2020a, katzFullyAutomatedEndtoend2022} can be readily implemented on GPU architectures.
Nevertheless, the EOB models necessitate the resolution of a set of ordinary differential equations, making their incorporation into parallel processing systems more complex.
To address the CPU computational challenges, data-driven surrogate models have been developed to significantly accelerate this process.
The surrogate models utilize machine learning methodologies, including the greedy algorithm \cite{fieldFastPredictionEvaluation2014} and proper orthogonal decomposition \cite{chatterjeeIntroductionProperOrthogonal2000, tiglioReducedOrderSurrogate2022}, to develop a reduced order model (ROM).  
These surrogate models for NR and EOB waveforms provide computationally efficient waveform representations while maintaining a high level of accuracy \cite{blackmanSurrogateModelGravitational2017, varmaSurrogateModelsPrecessing2019, islamEccentricBinaryBlack2021,yunSurrogateModelGravitational2021}.
\texttt{SEOBNRE\_S}\cite{yunSurrogateModelGravitational2021} and \texttt{NRSur2dq1Ecc} \cite{islamEccentricBinaryBlack2021} constructed the reduced-order surrogate model  for eccentric BBH waveform.

The rapid advancement of deep learning technologies in recent years has  led to extensive applications \cite{zhaoDawningNewEra2023} in GW detection \cite{georgeDeepLearningRealtime2018, georgeDeepNeuralNetworks2018,gabbardMatchingMatchedFiltering2018, zhaoDilatedConvolutionalNeural2024}, denoise\cite{wangWaveFormerTransformerbasedDenoising2024, zhaoSpacebasedGravitationalWave2023}, parameter estimation \cite{greenCompleteParameterInference2021a,daxRealTimeGravitationalWave2021, wangSamplingPriorKnowledge2022}, as well as in waveform generation \cite{chuaReducedorderModelingArtificial2019, khanGravitationalwaveSurrogateModels2021, nousiAutoencoderdrivenSpiralRepresentation2022, thomasAcceleratingMultimodalGravitational2022, fragkouliDeepResidualError2023,liaoDeepGenerativeModels2021, leeDeepLearningModel2021, khanInterpretableAIForecasting2022, shiCompactBinarySystems2024, yanModelingTimeEvolution2024a}. 
The surrogate model based on deep learning enables rapid EOB waveform generation on GPU architectures.
References \cite{chuaReducedorderModelingArtificial2019,khanGravitationalwaveSurrogateModels2021,nousiAutoencoderdrivenSpiralRepresentation2022, thomasAcceleratingMultimodalGravitational2022, fragkouliDeepResidualError2023} used an artificial neural network (ANN) to map GW parameters into their basis coefficients. 
However, these frameworks still rely on ROM techniques, which will be difficult to construct because they need a higher-order basis for the eccentricity waveform. 
Liao et al. \cite{liaoDeepGenerativeModels2021} employed a conditional autoencoder to learn waveforms' latent space. However, this method needs mapping parameters to latent space, resulting in some loss of accuracy.

\begin{figure*}
	\centering
	\includegraphics[width=0.9\linewidth]{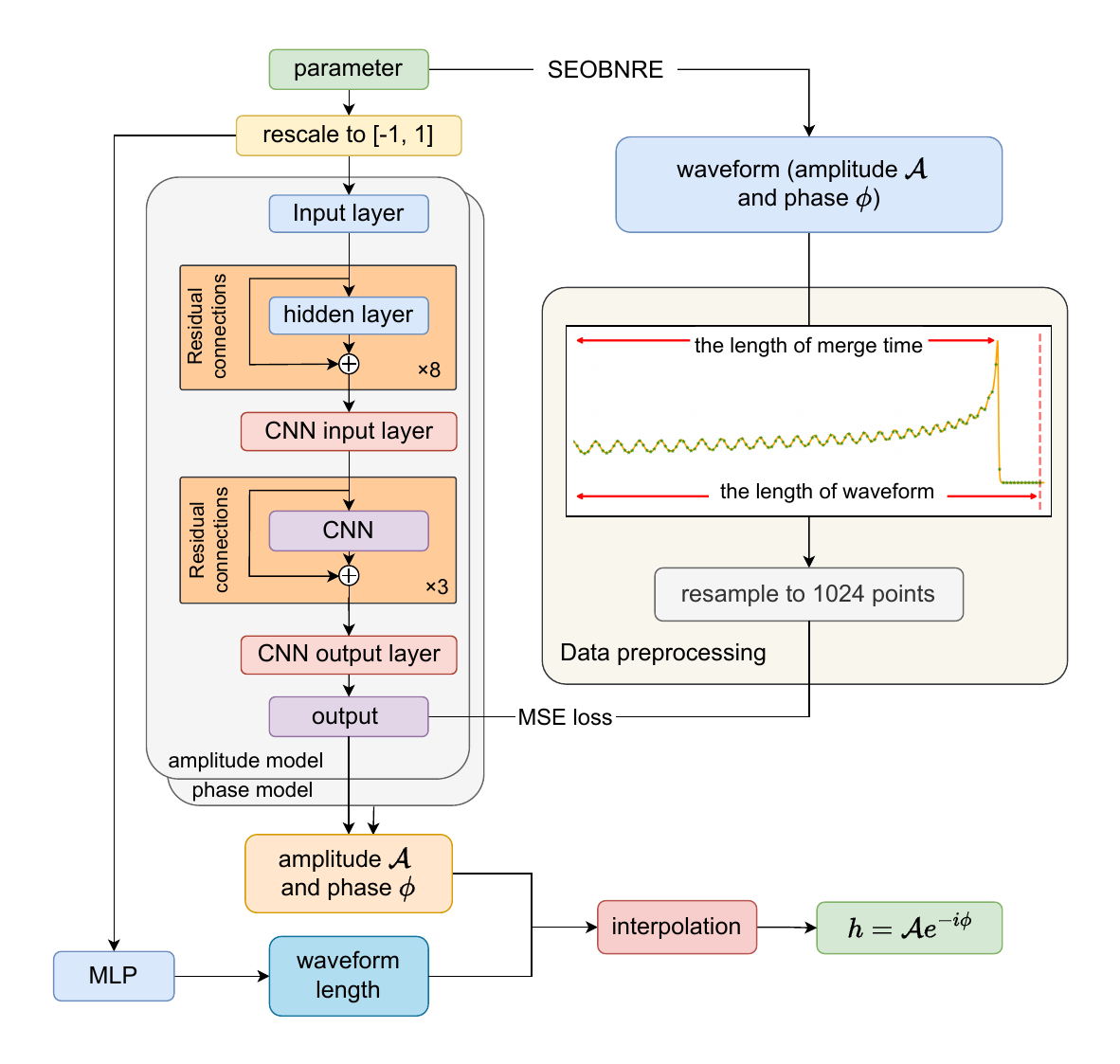}
	\caption{\textbf{The overview of data pre-processing and model architecture. }The data pre-processing section (right) illustrates the steps taken to prepare training samples, including waveform parameter normalization and length resampling. 
	}
	\label{fig:AI-model}
\end{figure*}

To accelerate the waveform generation for \texttt{SEOBNRE}, we propose a rapid generation framework \texttt{SEOBNRE\_AIq5e2} (Figure \ref{fig:AI-model}). 
In this framework, the waveforms are resampled to 1024 points during the training phase, and the model is trained on the resampled amplitude and phase. 
The resampled waveform is standardized to a fixed length of 1024 sampling points, which is unable to reconstruct to its original length directly. 
To address this issue, \texttt{SEOBNRE\_AIq5e2} uses a MLP to map waveform parameters to the waveform length, followed by an interpolation process to revert to the original waveform length.
Unlike other AI surrogate models, we do not need to employ the ROM method to generate low-rank basis representations of the waveform, our model maps waveform parameters to waveforms directly.
We utilize a single model to produce a shorter waveform that retains a substantial amount of the original waveform information.
The model can generate waveforms rapidly with the reference frequency $Mf=0.06$, mass ratio $q\in[1,5]$, eccentricity $e\in[0.0.2]$, and spin $\chi_{1,2}\in[-0.6,0.6]$, achieving an average mismatch of $1.02\times 10^{-3}$ and a signal waveform generation speed of 4.3 ms, 
which makes it possible for rapid parameter estimation and detection of the eccentricity BBHs.

The rest of this paper is organized as follows: Section~\ref{sec:data} and ~\ref{sec:met} describe data generation and \texttt{SEOBNRE\_AIq5e2} architecture, respectively. In Section~\ref{sec:res}, we present the \texttt{SEOBNRE\_AIq5e2} waveform accuracy and speed results and discuss the future development directions.
Finally, Section~\ref{sec:con} highlights our findings based on the results.

\section{Dataset construction}
\label{sec:data}

\subsection{Waveform generation}
The training dataset for this study was based on the \texttt{SEOBNRE} waveform model. \texttt{SEOBNRE} is an advanced model in the family of EOB approaches designed to incorporate features of NR simulations for eccentric, spinning binary systems with 2PN orbital radiation-reaction forces. It is particularly well-suited for accurately modeling the inspiral, merger, and ringdown phases of BBH coalescences, including systems with non-negligible eccentricity and spins \cite{caoWaveformModelEccentric2017,liuHighermultipoleGravitationalWaveform2022, liuEffectiveonebodyNumericalrelativityWaveform2024}.

Each waveform was generated in the form of the two polarizations, $h_+$  and $h_{\times}$, over a time interval that captures the full inspiral, merger, and ringdown. 
To simplify model training, we separate the learning of the waveform's amplitude and phase components:
\begin{equation}
\begin{aligned}
    h(t,\lambda) &= h_+(t, \lambda) - ih_{\times}(t, \lambda) \\
    &=  \mathcal{A}(t, \lambda) e^{ -i\phi(t, \lambda)},
\end{aligned}
\label{equ:wavefrom}
\end{equation}
where $\mathcal{A}(t,\lambda)$ represents the waveform amplitude, $\phi(t,\lambda)$ is the waveform phase, and $\lambda$ represents the GW source parameter. 
By decomposing the waveform in this way, each component is processed independently, allowing the model to focus on specific features of the amplitude and phase without interference from one another. 
The parameter space of the train and test dataset is shown in
Table \ref{tab:1}. In this study, we restrict our analysis to the dominant $(2, 2)$ mode of the GW signal.

During dataset generation, the dimensionless reference frequency was fixed at $Mf=0.06$, sample rate is set to 4096Hz, and the total mass of the BBH was set to $60M_{\odot}$ ($f_0\approx 20\text{Hz}$). 
The value of $Mf$ is fixed, and it can be rescaled across different total masses.
This guarantees that the generated waveforms span the frequency bands of the ground-based GW detectors and the temporal length appropriate for the detection and analysis of stellar.

\begin{table*}
\caption{Parameter distribution of training dataset and test dataset}
\label{tab:1}
\centering
    \begin{tabular}{p{2cm}p{8cm}p{4cm}}
        \toprule
        Parameter    & Description\centering & Parameter distribution \\
        \midrule
        $M_{tot}$   & Total mass of BBHs $m_1 + m_2$   & Fix at $60 M_{\odot}$  \\
        $q$         & Mass ratio $\frac{m_1}{m_2}$ and $q\geq 1$           &  Uniform $[1,5]$  \\
        $e_0$         & eccentricity at $f_0$           &  Uniform $[0,0.2]$  \\
        $\chi_1, \chi_2$  & Spin parameters of two black holes      &  Uniform $[-0.6, 0.6]$  \\
        \bottomrule
    \end{tabular}
  \end{table*}

\subsection{Merger time alignment}

\begin{figure*}
    \centering
    \subfigure[Merger time alignment, merge time is at zero point.]{
        \label{fig:high_freqs1}
        \includegraphics[width=0.45\linewidth]{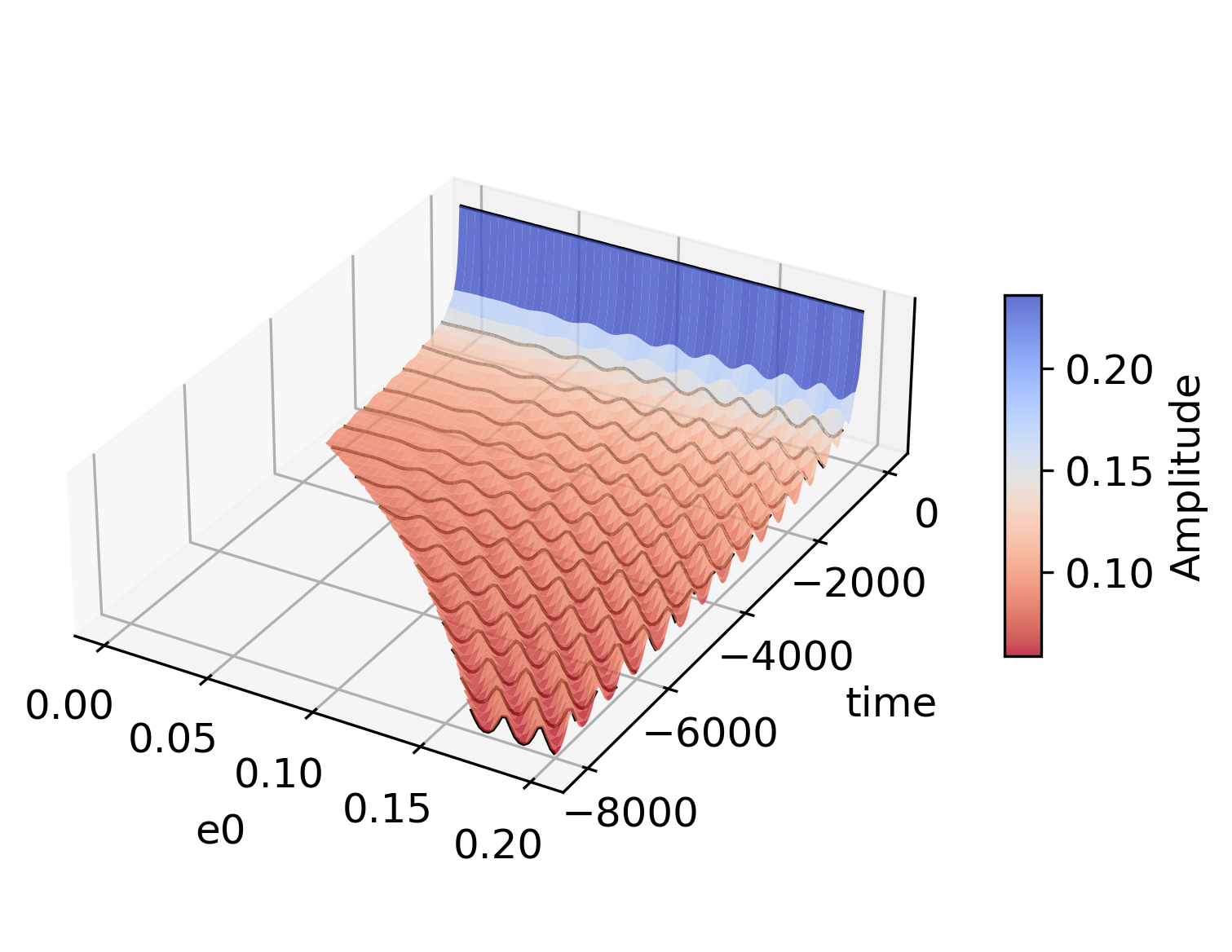}
    }
    \subfigure[Waveform resampling, initial evolution time is at zero point.]{
        \label{fig:high_freqs2}
        \includegraphics[width=0.45\linewidth]{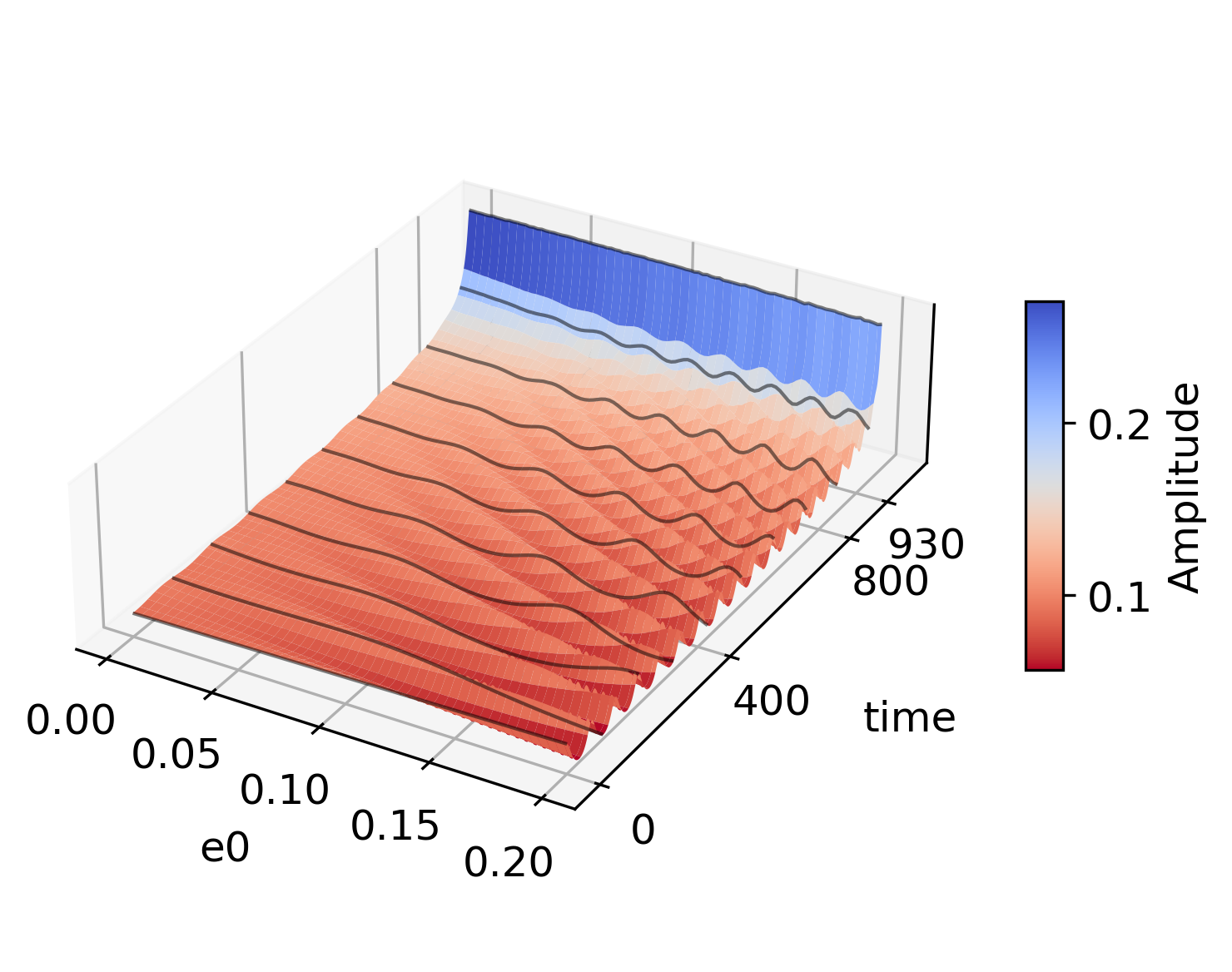}
    }
    \caption{\textbf{
    The surface of $\mathcal{A}(e_0;t)$ for the (2,2) mode.} Here we set $q=2.2$ and $\chi_{1,2} = 0$. The black lines that cover the surface indicate the fluctuation of the amplitude $\mathcal{A}_t(e_0)$ in the parameter space at a fixed time point, and the time axis is the corresponding sampling point. (a) shows fluctuations in the parameter space $e_0$. (b). waveforms are resampled to same length to ensure uniformity and crucial information preservation. }
    \label{fig:high_freqs}
\end{figure*}

Previous approaches constructed a data-driven surrogate model by aligning merging times. 
The eccentric GW waveform has a far more complex waveform than the non-ecc waveform, and waveforms with varied parameters exhibit apparent fluctuations. Figure \ref{fig:high_freqs1} shows fluctuations in the parameter space $e_0-t$.
Unfortunately, data-driven algorithms prefer to learn smooth components, and these fluctuation components are more difficult to fit. 
For example, the \texttt{SEOBNRE\_S} model employs over 300 basic functions to reach an accuracy level comparable to that achieved with only 100 bases in non-eccentric waveform models \texttt{NRSur7dq4} at 7 dimension parameters \cite{varmaSurrogateModelsPrecessing2019}. 

At the same time, given a fixed-length output structure that uses a non-recurrent neural network architecture, the shorter waveforms of the model require zero-padding to match the output length. 
This approach introduces an imbalance in the training sample distribution due to inconsistencies in waveform length across different parameters. 
If the waveform is cut to the smallest length in the training set, the gravitational wave information is lost. 
Furthermore, neither the zero-padding nor waveform truncation can alleviate the fluctuation phenomena in parameter space.


We employ an adaptive resampling technique to address sample imbalance and reduce information loss from truncation. This approach resamples waveforms of different lengths to a consistent number of sampling points (Figure \ref{fig:high_freqs2}), ensuring consistency across samples while preserving essential information with minimal loss. 
Moreover, the fluctuation in parameter space is alleviated in the early sample points. 
Here we use cubic spline interpolation for resampling.
In section \ref{subsec:resample} we will introduce how to resample the waveform.

\subsection{Waveform resample and interpolation}
\label{subsec:resample}
To standardize the length of waveforms across different parameters, each waveform was rescaled to a length corresponding to 1.1 times from the $t_0$ to the merger time $t_m$, which is called the original waveform duration $t_{\text{orig}}$:
\begin{equation}
    t_{\text{orig}} = 1.1\times(t_m - t_0).
\end{equation}
Because the sample rate is constant at 4096 Hz, for simplicity, $ t_{\text{orig}}$ is an integer, referring to the number of sampling points.

Subsequently, each waveform is resampled to 1024 points using cubic spline interpolation, which ensures consistent data representation across the dataset and preserves the key features of the signal while reducing the computational complexity. 
The resampling points are:
\begin{equation}
    t_{\text{resample}} = \frac{i}{N-1} \cdot (t_{\text{orig}} - 1), \quad i = 0, 1, \dots, 1023,
    \label{equ:resample}
\end{equation}
where $N$ refers to the length after resample, and we set $N=1024$.

The model will generate a sequence of 1024 points, which must be returned to the original time series. To recover the original length, the model will output the expected $\hat{t}_{\text{orig}}$ and use cubic spline interpolation to the recovered time series $t_{\text{interp}}$:
\begin{equation}
    t_{\text{interp}} = \frac{i}{\hat{t}_{\text{orig}}-1} \cdot 1023, \quad i = 0, 1, \dots, \hat{t}_{\text{orig}}-1.
\end{equation}

This approach also encounters high-frequency fluctuations within the parameter space before the merger time; however, in Figure \ref{fig:high_freqs2} these fluctuations are significantly mitigated during the early stages of evolution. 
This reduction enables the model to efficiently learn the overall structure of the waveform and capture the initial signal characteristics. 
With further training to incorporate higher-frequency details, the model achieves an output of high-accuracy waveforms.
Figure \ref{fig:AI-model} illustrates the pre-processed waveforms.
This resampling approach guarantees efficient training of the AI model by maintaining a uniform input size. 

Our model produces two time series outputs: amplitude and phase, along with the corresponding waveform duration. 
The model output time series are interpolated to the defined waveform duration, ensuring consistency with the physical duration of the waveform. The duration of the final output waveform varies depending on the input waveform parameters.

The parameters of the training set and the test set are randomly sampled in the parameter space.
A total of 500,000 waveforms were generated for the training set, while the additional 10,000 waveforms were generated for the test set. 

\section{SEOBNRE\_AIq5e2 model}
\label{sec:met}
The \texttt{SEOBNRE\_AIq5e2} model inputs the GW source parameter and outputs the generated waveform that is derived from three characteristics: the amplitude, phase, and length of the waveform.
This structured approach enables a comprehensive representation of the GW, facilitating accurate modeling of its characteristics based on the input source parameters.

\subsection{Amplitude and phase generation model}
The \texttt{SEOBNRE\_AIq5e2} model combines a multi-layer perception (MLP) with a convolutional neural network (CNN). 
This hybrid approach leverages both the MLP's capability to learn complex parameter-space mappings and the CNN's efficiency in modeling temporal correlations.
Respective models are trained for amplitude $\mathcal{A}(t,\lambda)$ and phase $\phi(t,\lambda)$ prediction. 

\textit{Normalizing the input parameters.\textemdash} Since the value ranges between different parameters are inconsistent, a linear mapping is applied to each parameter to standardize the input. Specifically, each parameter $p$ is linearly mapped to the range $[-1, 1]$ using the following transformation:
\begin{equation}
    p_{\text{norm}} = 2 \times \frac{p - p_{\text{min}}}{p_{\text{max}} - p_{\text{min}}} - 1,
\end{equation}
where $p_{\text{min}}$  and $p_{\text{max}}$ represent the upper and lower limits of the parameter range of parameter $p$ respectively. 
This normalization ensures that all input parameters are on a consistent scale, enabling a more stable and efficient training of the model.

\textit{MLP.\textemdash} The MLP layer used in this model is designed to map input parameters to high-dimensional latent representations. 
It facilitates the model to capture complex relationships between input parameters and waveform outputs. 
The MLP consists of an input layer, followed by 8 hidden layers, each with residual connections and ReLU activation.

The input to the MLP is a 4-dimensional vector that represents the parameters of the wave source. The input layer maps this 4-dimensional input into a 1024-dimensional hidden space. The transformation applied in the input layer can be represented as:
\begin{equation}
    \mathbf{h}^{(0)} = \text{ReLU}\left( \mathbf{W}^{(0)} \mathbf{p} + \mathbf{b}^{(0)} \right),
\end{equation}
where $\mathbf{p} \in \mathbb{R}^4$ is the input vector, $\mathbf{W}^{(0)} \in \mathbb{R}^{1024 \times 4}$  and  $\mathbf{b}^{(0)} \in \mathbb{R}^{1024}$ are the weight matrix and bias vector of the input layer, respectively. 
Following the input layer, each hidden layer has a hidden size of 1024. Residual connections are incorporated into each hidden layer to stabilize gradient flow and model performance. For each hidden layer $l$, where $l = 1, \dots, 8$, the output is computed as follows:
\begin{equation}
    \mathbf{h}^{(l)} = \text{ReLU}\left(  \mathbf{W}^{(l)} \mathbf{h}^{(l-1)} + \mathbf{b}^{(l)} \right)  + \mathbf{h}^{(l-1)},
\end{equation}
where $\mathbf{W}^{(l)} \in \mathbb{R}^{1024 \times 1024}$, $\mathbf{b}^{(l)} \in \mathbb{R}^{1024}$ are the bias vector for the $l$-th hidden layer, $\mathbf{h}^{(l-1)} \in \mathbb{R}^{1024}$ is the output of the previous layer, and the residual connection adds this to the transformed output of the current layer. 

\textit{CNN.\textemdash} The output of the MLP is subsequently inserted into the CNN.
Firstly, the MLP output is transformed into a 32-channel representation. 
This multi-channel expansion enhances the model's capacity to capture intricate waveform features. 
Subsequently, three consecutive convolutional layers with residual connection are employed to smooth the waveform across channels, refining the feature representation and preserving essential temporal patterns. 
Finally, an additional convolutional layer condenses the 32-channel output back into a single channel, yielding a processed waveform that retains both the structural integrity and details for high-accuracy waveform reconstruction.
This hierarchical approach allows the CNN to learn intricate temporal dynamics associated with the gravitational waveforms, enhancing the model's generation accuracy.

Additionally, each convolutional layer utilizes the ReLU activation function and instance normalization (InstanceNorm) to promote stable training and enhance feature extraction across the network.
The InstanceNorm \cite{ulyanovInstanceNormalizationMissing2017} is computed as:
\begin{equation}
    \hat{\mathbf{h}}^{(l)} = \frac{\mathbf{h}^{(l)} - \mu}{\sigma} + \beta,
\end{equation}
where $\mu$ and $\sigma$ are the mean and standard deviation of the features for each instance, respectively. 
This normalization helps to reduce covariate shift during training, leading to faster convergence and improved model stability.

\subsection{Waveform-length model}
To address the challenge of varying waveform lengths across different parameter configurations, we resample all waveforms at the same length. 
However, this standardization renders the actual waveform length unknown for each case. To predict the original waveform length, we employ a 4-layer MLP model with a hidden layer size of 1024. 
This MLP model is trained separately to estimate the waveform length based on the waveform parameter, enabling the framework to subsequently apply interpolation. 
Specifically, we use the cubic spline interpolation, a method known for its smoothness and efficiency in approximating continuous functions, to recover the waveform.
This approach ensures consistency in input dimensions while retaining essential information about the waveform’s natural length. 


\subsection{Training and Verification}\label{subsec:tni}
During training, the Adam optimizer \cite{Kingma2017} is used to optimize the model. We use signal RTX 4090 GPU to train the amplitude, phase and waveform length model in this study for approximately 12 hours, respectively. 
The learning rate was initialized at $10^{-4}$ and decayed by a factor of 0.95 every 50 epochs to gradually reduce the step size and enhance convergence stability over the course of training.

To assess the MLP-CNN model's capacity to generate GW  waveform, we measure its recovery ability by comparing \texttt{SEOBNRE\_AIq5e2} waveform with SEOBNRE waveform.
Eq. \ref{equ:overlap} calculates the overlap $\mathcal{O}$ between the target waveform and the generated waveform using the \texttt{pycbc} library \cite{alex_nitz_2024_10473621}. The overlap value ranges from 0 to 1, with higher values indicating greater similarity between the generated waveform and the target waveform.
    \begin{equation}
        \mathcal{O}({h}_1, {h}_2)=
        \max_{t_c, \phi_c}\left(\hat{h}_1|\hat{h}_2\right)^{1/2} ,
        \label{equ:overlap}
    \end{equation}
    \begin{equation}
        \hat{h} = \frac{h}{\sqrt{(h|h)}},
    \end{equation}
where $(a|b)$ represents the inner product:
    \begin{equation}
        (h_1|h_2) =4\text{Re}\int_{f_{\min}}^{f_{\max}}\frac{\tilde{h}^*_1(f)\tilde{h}_2(f)}{S_n(f)}df, \\
            \label{equ:inner}
    \end{equation}
where $S_n$ is one-sided power spectral density (PSD) of the detector noise and is set to $1$. Finally, the mismatch between two waveform is defined as
\begin{equation}
    \mathcal{M}({h}_{1}, {h}_2) = 1-\mathcal{O}({h}_1, {h}_2)
    \label{equ:mismatch}
\end{equation}

\section{Results}
\label{sec:res}
In this section, we evaluate the model’s performance based on two primary metrics: waveform mismatch and generation speed. Mismatch is assessed to quantify the accuracy of \texttt{SEOBNRE\_AIq5e2} waveforms relative to the target waveforms, providing insights into accuracy across a range of parameter configurations. 
Generation speed is examined to demonstrate the efficiency of the \texttt{SEOBNRE\_AIq5e2} waveform model.

\subsection{Mismatch Reuslts}

\begin{figure}
    \centering
    \includegraphics[width=1\linewidth]{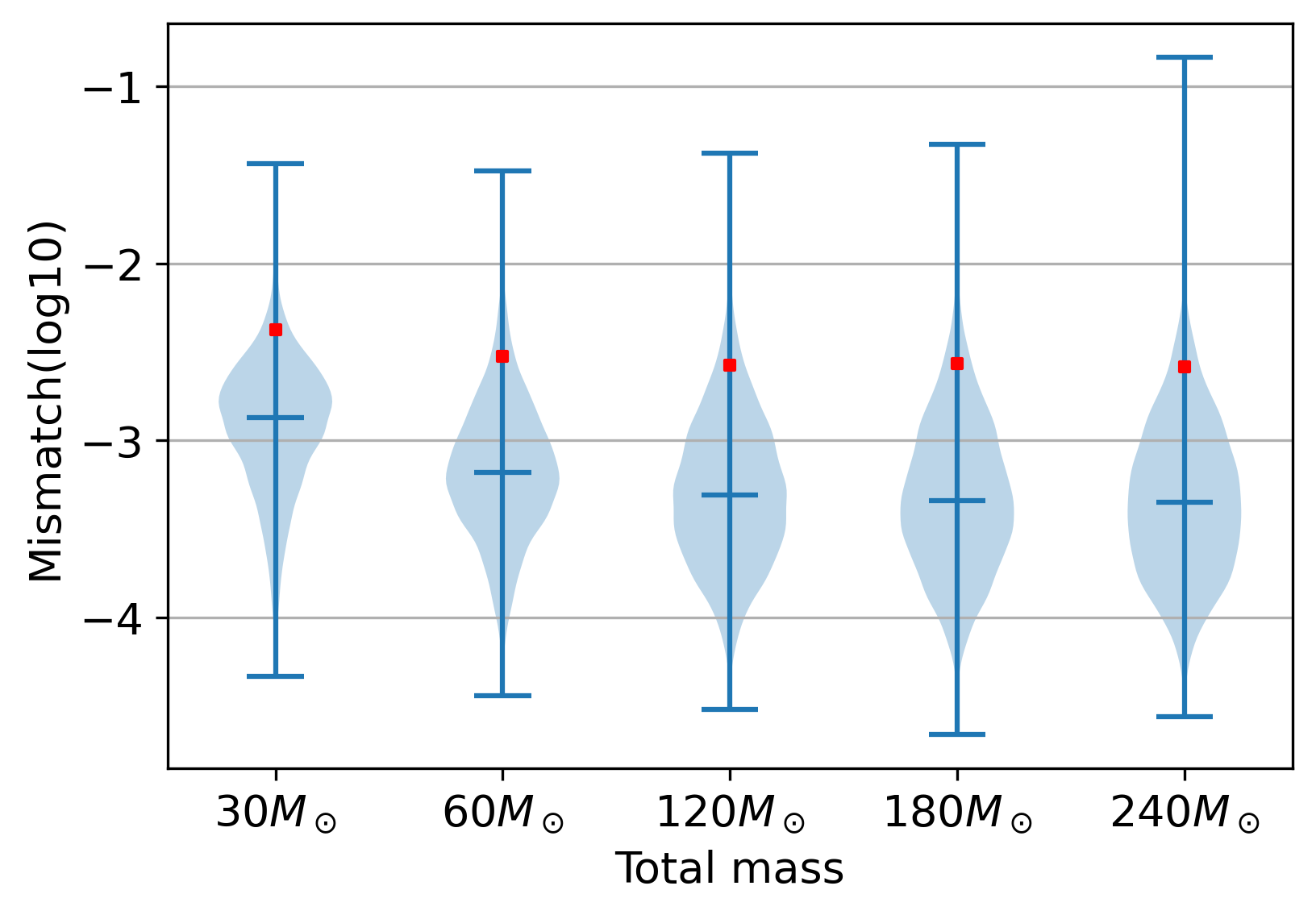}
    \caption{\textbf{Distribution of mismatches between \texttt{SEOBNRE} and \texttt{SEOBNRE\_AIq5e2} on the test set.} During test we rescale the waveforms of $M_t=60M_\odot$ to different total masses. The red mark represents the 95\% percentile of the mismatch.}
    \label{fig:overall_mismatch}
\end{figure}

\begin{figure*}
    \centering
    \subfigure{
        \label{fig:showcase2}
        \includegraphics[width=0.8\linewidth]{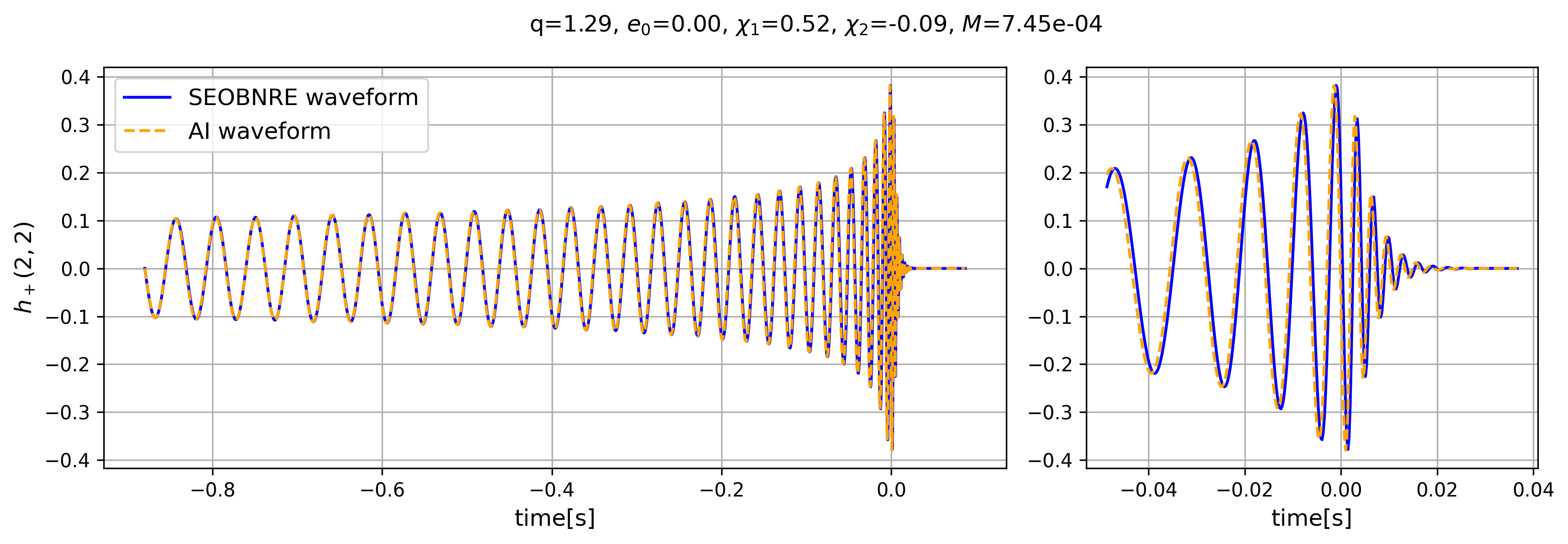}
    }
    \subfigure{
        \label{fig:showcase1}
        \includegraphics[width=0.8\linewidth]{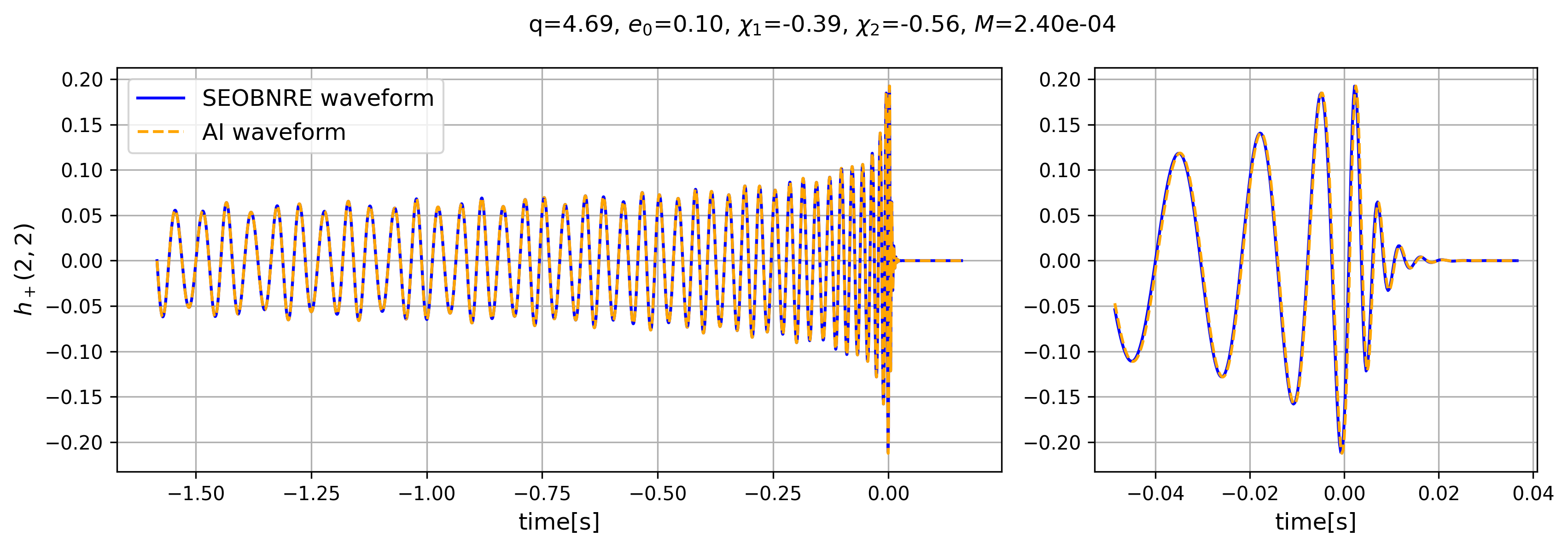}
    }
    \subfigure{
        \label{fig:showcase3}
        \includegraphics[width=0.8\linewidth]{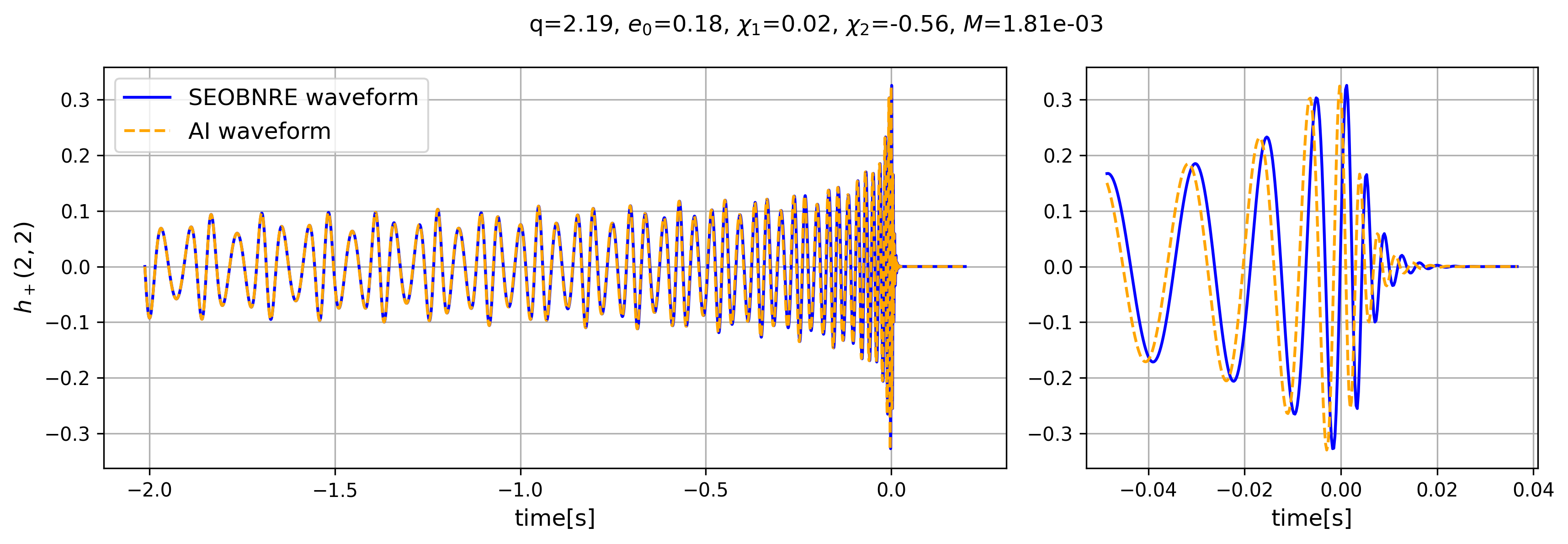}
    }
    \caption{\textbf{Comparison of generated waveforms between \texttt{SEOBNRE\_AIq5e2} and \texttt{SEOBNRE}.} The figure demonstrates the waveform generation capability of \texttt{SEOBNRE\_AIq5e2} model across a wide range of parameter configurations, where $M$ stands for mismatch value.}
    \label{fig:showcase}
\end{figure*}

Figure \ref{fig:overall_mismatch} shows the mismatch between the targets and generated waveforms of \texttt{SEOBNRE\_AIq5e2} at different total masses. 
In the test set ($M_t=60M_\odot$), the mean mismatch is $1.02 \times 10^{-3}$, with mismatch values ranging from a minimum of $3.63 \times 10^{-5}$ to a maximum of $3.31 \times 10^{-2}$. 
Additionally, Figure \ref{fig:showcase} illustrates showcases of waveform generation by our AI model, demonstrating its strong capability to replicate target waveforms across wide GW parameter range accurately.

Using a reference total mass of $M_t = 60 M_\odot$, we evaluate the mismatch for various total masses, specifically $M_t = (30, 60, 120, 180, 240) M_\odot$ and the violin plots are shown in Figure \ref{fig:overall_mismatch}.
The mean mismatch remains relatively stable and low across different mass scalings, with a slight improvement in mismatch observed at higher total masses compared to lower total masses. 
We observe that the maximum mismatch value increases when total masses become large. 
This tendency occurs as higher-mass systems are more sensitive to the merger stage \cite{khanGravitationalwaveSurrogateModels2021}, where any tiny errors in generation waveforms can be amplified.

\begin{figure*}
    \centering
    \includegraphics[width=0.8\linewidth]{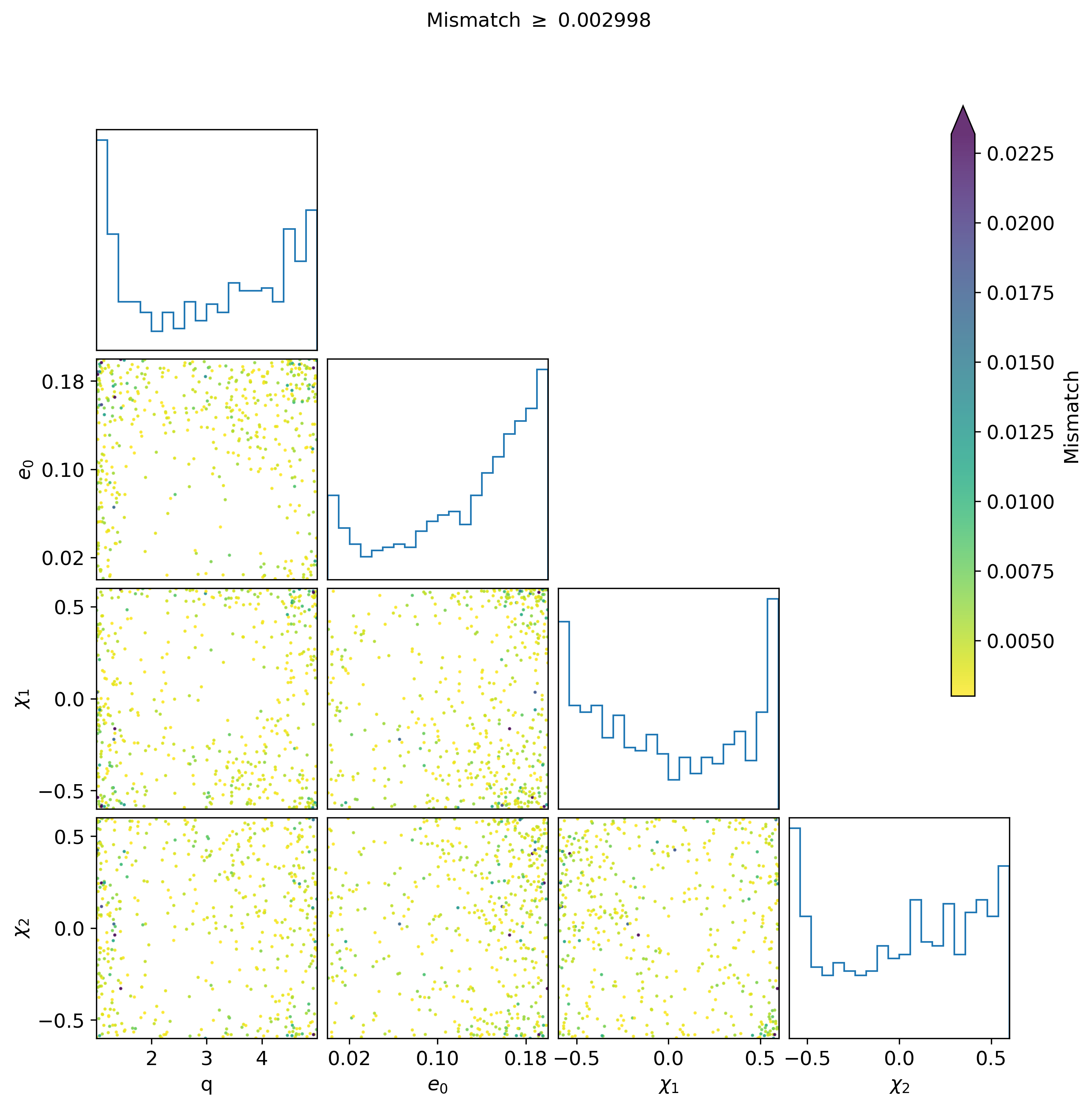}
    \caption{\textbf{Distribution of mismatches between target waveforms and AI-generated waveforms on the test set}. Here is a scatter plot of mismatch above the 95\% quantile at $M_t=60M_\odot$ in the (q, $e_0$, $\chi_1$, $\chi_2$) parameter space. Higher mass ratio, eccentricity, and spin all exhibit higher mismatch values. In addition, the area with a low mass ratio will have a larger mismatch value. The other total masses are similar to the $M_t=60M_\odot$.}
    \label{fig:overall_mis_scatter}
\end{figure*}

Figure \ref{fig:overall_mis_scatter} displays the distribution in parameter space of waveforms with mismatch values exceeding the 95th percentile at $M_t=60M_{\odot}$. 
The histogram of Figure \ref{fig:overall_mis_scatter} shows that the mismatch is larger where the mass ratio is higher or lower. Simultaneously, the higher the eccentricity $e_0$ and spin $\chi_{1,2}$, the bigger the mismatch.
While the scatter plot is generally uniformly distributed, there are some tendencies for clustering in certain regions. 
In the mismatch scatter plot across the $q$-$e_0$ parameter space, regions of higher mismatch are predominantly located at larger eccentricities, larger or smaller mass ratios. 
Similarly, there is a cluster in high-spin areas. 
A similar pattern of distribution is observed even when the waveforms are rescaled to different total masses. 
As the initial eccentricity and mass ratio increase, the waveform duration extends, adding complexity to model training and requiring a greater interpolation factor, both of which contribute to higher mismatch values.

\subsection{Waveform generation speed}
Compared to EOB waveform models, a significant advantage of data-driven waveforms lies in their substantially faster generation speed. 
On a single RTX 4090 GPU, the model achieves a waveform generation speed of 4.3 ms (Table \ref{tab:2}). 
\texttt{SEOBNRE\_AIq5e2} is also competitive in terms of frequency domain waveform generation speed and can provide a more comprehensive eccentric waveform.
There is nearly $500\times$ faster than \texttt{SEOBNRE} waveform generation at one CPU core (Intel 2.3GHz Xeon 8336C). 
The model demonstrates efficient training speeds. With a training set of 50,000 samples, the time required to train a single model for amplitude, phase, or waveform length is approximately 12 hours over 1,500 epochs. 

Another advantage of using GPUs is their high degree of parallelism. 
In Table \ref{tab:3}, we compared the generation speed of the single GPU and the 64-core CPUs (2$\times$Intel 2.3GHz Xeon 8336C) with batch=(1, 10, 100,  1000). 
To benchmark generation speed against \texttt{SEOBNRE}, we employ multiple processes for waveform generation, with the batch size in Table \ref{tab:3} representing the number of concurrent processes used. 
Even at larger batch sizes, the model has remarkable acceleration performance, which is a great benefit for Bayesian inference.
For example, \texttt{SEOBNRE\_AIq5e2} generates the required 1000 waveforms in 0.812 seconds using parallel-tempered MCMC tools \cite{littenbergBayesianApproachDetection2009a}, whereas 64-core CPUs take around 39.34 s. 
This means \texttt{SEOBNRE\_AIq5e2} is about 48$\times$ faster when run in highly parallel setups.
The AI model generates waveforms at a consistent speed across different parameters and efficiently handles longer waveforms in extreme cases. 
Although the interpolation process becomes slower with increasing waveform length, the time required for interpolation constitutes only 1/10 to 1/100 of the total time that is needed to generate the AI-produced part. 
For the case of generating a large number of waveforms in parallel, the efficiency of performing parallel interpolation calculations can be further improved on this basis (Table \ref{tab:3}).

\begin{table}
\caption{\textbf{Time that required to generate a single waveform.} Here, we choose the average generation speed of 100 waveforms. The waveform production settings are the same as in Section \ref{sec:data}. For the frequency domain waveform, we set $f_{min} = 20Hz$ and $\Delta f = 1/8Hz$.}
\label{tab:2}
\centering
\begin{threeparttable}
    \begin{tabular}{p{3.5cm}<{\centering}p{3.5cm}<{\centering}}
        \toprule
        waveform    & speed  \\
        \midrule
        *\texttt{SEOBNRE\_AIq5e2}(GPU) & 4.288 ms  \\
        *\texttt{SEOBNRE\_AIq5e2}(CPU) & 9.856 ms  \\
        *\texttt{SEOBNRE}        &  2.133 s  \\
        *\texttt{EccentricTD}    & 61.948 ms  \\
        **\texttt{TaylorF2Ecc}    & 0.238 ms  \\
        **\texttt{EccentricFD}    & 2.465 ms  \\
        \bottomrule
    \end{tabular}
    \begin{tablenotes}
\item[*] Time domain waveform
\item[**] Frequency domain waveform
\end{tablenotes}
\end{threeparttable}
\end{table}

\begin{table*}
\caption{
\textbf{The computational efficiency of \texttt{SEOBNRE\_AIq5e2} under different batch sizes.} We evaluated the generation speed of \texttt{SEOBNRE\_AIq5e2} on both GPU and CPUs, and \texttt{SEOBNRE} generation speed across varying process counts on the 64-core CPUs.
}
\label{tab:3}
\centering
\begin{threeparttable}
    \begin{tabular}{p{2cm}<{\centering}p{2cm}<{\centering}p{2cm}<{\centering}p{2cm}<{\centering}p{3cm}<{\centering}}
        \toprule
        Batch size   & GPU & CPU & \texttt{SEOBNRE} & \thead{speed up\\(\texttt{SEOBNRE}/GPU)}\\
        \midrule
        $1$    & 4.29e-3 s & 9.86e-3 s & 2.133 s  & 497\\
        $10$   & 0.0123 s  & 0.0277 s  & 2.824 s  & 229\\
        $100$ & 0.0819 s  & 0.2170 s  & 4.514 s  & 55\\
        $1000$ & 0.8123 s  & 1.2404 s  & 39.340 s & 48\\
        \bottomrule
    \end{tabular}
\end{threeparttable}
\end{table*}


\section{Conclusion and Discussion}
\label{sec:con}
In this study, we present the \texttt{SEOBNRE\_AIq5e2} model, establishing a novel framework for waveform surrogate modeling based on deep learning. 
This is the first model to use deep learning in the field of eccentric waveform generation. 
Through this approach, \texttt{SEOBNRE\_AIq5e2} advances the efficiency and accuracy of eccentric waveform generation, offering a powerful tool for enhancing gravitational wave data analysis. 
At the same time, \texttt{SEOBNRE\_AIq5e2} is readily portable to the CPU, and the generation speed of a 64-core CPU is merely twice as a GPU. This implies that users without a GPU may assure their generating efficiency under the CPU.

In contrast to existing surrogate methods, this paper introduces a resampling-interpolation approach to enable variable-length waveform generation within a data-driven framework. 
This approach effectively addresses the frequency principle limitation \cite{xuFrequencyPrincipleFourier2020} in deep learning, which significantly reduces oscillations in the parameter space. 
\texttt{SEOBNRE\_AIq5e2} achieves a mean mismatch of $1.02 \times 10^{-3}$ under conditions of $Mf = 0.06$, $1 < q < 5$, $0 < e < 0.2$, and spin parameters $-0.6 < \chi_{1,2} < 0.6$, demonstrating its robustness and accuracy across diverse waveform conditions.
The model demonstrates robust accuracy across varying total mass, and we have shown that cubic spline interpolation effectively maintains the precision of waveform generation.

Compared to the \texttt{SEOBNRE} waveform, \texttt{SEOBNRE\_AIq5e2} model achieves significantly faster generation speeds, requiring only 4.3 ms to produce a single waveform, with an improvement of two orders of magnitude over SEOBNRE. 
Leveraging GPU-based high-performance computing, the model enables simultaneous generation of multiple waveforms, reducing computational costs substantially and enhancing efficiency in large-scale waveform synthesis.
Methods such as importance sampling, density estimation, and PTMCMC require the parallel generation of numerous waveforms for each evaluation, highlighting the advantages of GPU high-performance computing. 
The \texttt{SEOBNRE\_AIq5e2} model is well-suited to these demands, as it significantly reduces waveform generation time. 
This model has the potential to greatly shorten parameter estimation times and enhance the efficiency of parameter estimation for eccentricity waveforms.


In future work, we aim to broaden the parameter space and incorporate higher modes within our model to enhance its applicability for the detection and parameter estimation of BBHs in eccentric orbits. Additionally, we will explore the potential of fine-tuning the \texttt{SEOBNRE\_AIq5e2} model using NR waveforms to further improve accuracy. 
We will also look at using AI-driven interpolation approaches to enable a completely differentiable waveform template, allowing us to leverage gradient-based MCMC techniques like Hamiltonian Monte Carlo \cite{bouffanaisBayesianInferenceBinary2019} to estimate parameters more efficiently.

\section*{Acknowledgments}
	This research was supported by the National Key Research and Development Program of China Grant No. 2021YFC2203001 and in part by the NSFC (No. 11920101003 and No. 12021003). 
	Z.C. was supported by the “Interdisciplinary Research Funds of Beijing Normal University" and CAS Project for Young Scientists in Basic Research YSBR-006.
      This work was also supported in part by the Peng Cheng Laboratory and Peng Cheng Cloud-Brain.

\bibliographystyle{myREVTeX4-2}
\bibliography{references}
\end{document}